\documentclass[3p,times]{elsarticle}

\usepackage{ecrc}

\volume{00}

\firstpage{1}

\journalname{Nuclear Physics A}

\runauth{}

\jid{nupha}

\usepackage{amssymb}

\usepackage[figuresright]{rotating}

\begin{document}

\begin{frontmatter}



\dochead{}

\title{Theory summary. Hard Probes 2012}


\author{Carlos A. Salgado}

\address{Departamento de F\'\i sica de Part\'\i culas and IGFAE\\Universidade de Santiago de Compostela. E-15782 Santiago de Compostela (Galicia-Spain)}
\address{Physics Department, Theory Unit, CERN, CH-1211 Gen\`eve (Switzerland)}

\begin{abstract}
I provide a summary of the theoretical talks in Hard Probes 2012 together with some personal thoughts about the present and the future of the field.
\end{abstract}


\end{frontmatter}



The second nuclear run at the LHC, in 2011, produced fifteen times more luminosity than the first one in 2010. This higher precision  data, first showed in this conference, marks the beginning of the new studies of hard probes of the hot and dense matter produced in heavy-ion collisions. The precision of the data is now often smaller than the theoretical uncertainties, reaching higher virtualities where the use of perturbative QCD techniques are under better control. New observables as reconstructed jets, or more systematics as the case of the different quarkonium states, call for new theoretical developments. Some of these developments started to be developed in the last years and are at a stage of being usable to phenomenological studies: new understanding of the in-medium parton shower; Monte Carlo codes; developments in quarkonia in-medium in lattice and effective theories or potential models; the knowledge of the initial conditions and the role of saturation, etc. In any of these fields, the theoretical progress in the last years, as reported in this conference, is very remarkable. Based on these rapid developments, an understanding of the experimental data of the first LHC running period, as well as the corresponding one from RHIC, based on solid calculations can be expected before the next LHC running period starts. 

Here I review the theory talks presented in the Hard Probes 2012 conference in Sardinia, Italy, often providing my own point of view and where I think that progress in the coming years is both needed and more promising. The different sections are rather obvious, and, most frequently follow the different topics in the scientific program. In the list of references I only quote the talks in this conference, the relevant references can be found in the corresponding contributions.

\section{Initial State}
\label{initialstate}

A precise knowledge of the partonic wave function of the nucleus is recognized to be a prerequisite for precise phenomenological studies of hot and dense matter. One of the main questions is the role played by non-linear terms in the evolution equations of the partonic distributions, or, in other words, when these terms are needed. This is also a fundamental question in QCD which connects to the asymptotic behavior where unitarity corrections are expected to play a central role. During the last twenty years, the original McLerran-Venugopalan (MV) model, in which the gluon distributions are computed in a semiclassical approach has evolved into a general framework, the Color Glass Condensate (CGC), to include the different physics from NLO non-linear evolution equations to thermalization, with different degrees of progress depending on the complexity of the task. 

One of the most urgent and fundamental questions is the description of the nuclear partonic structure. As for the case of the proton, these studies need of experimental data to fix some of the unknowns --- boundary or initial conditions in the evolution equations. The absence of precise and abundant DIS data has been historically the main limitation. In these conditions, a reasonable approach is to first study the proton structure, where abundant data exists, and then to compare with the nuclear data to extract {\it nuclear effects}. This is the traditional approach in nuclear PDF analyses, whose present status was reviewed in this conference by Kari Eskola \cite{here:eskola}. This is also the approach that is being followed more recently in the CGC framework by extensively using the best theoretical tools at our disposal, the NLO Balitsky-Kovchegov equations: fits to proton DIS provide a very good description of experimental data which can then be used as reference for the nuclear case. This allows for precision analyses of the data, as presented in the conference by Paloma Quiroga \cite{here:quiroga}, which promise to be the best tool to define the position in phase space of the saturation scale. One of the main problems in these approaches, more serious in the nuclear case, is the role of the impact parameter: the long range Coulomb tails in the transverse distributions of dipoles during the BFKL-like evolution, make the dilute edges of the proton or nucleus to grow too fast. A theoretical solution of this problem does not seem to appear soon, as it deals with the long-distance, non-perturbative, part of the interaction. The most successful phenomenological solution consists in solve locally, in transverse plane, the BK evolution equations in a Monte Carlo approach which fixes the local density from Glauber profiles. This approach is probably the most successful one in reproducing the multiplicities as a function of centrality in PbPb collisions at the LHC, providing also a good description at RHIC with basically one free parameter, once the proton distributions are known. When used to compute transverse momentum differential spectra, however, the uncertainties in the proton distributions translate, e.g. into large differences in the nuclear modification factor $R_{pPb}$ at the LHC, while still providing a reasonable description at RHIC. The origin of these uncertainties is the value of the anomalous dimension in the transverse momentum dependence of the unintegrated gluon distributions. This makes the approach to have limited predicting power, especially close to central rapidities, while the situation is better at large rapidities. The next proton-nucleus run at the LHC will provide the needed constraints to these models. These and other issues where reviewed by Javier Albacete \cite{here:albacete}.

Within the DGLAP approach, Ilkka Helenius presented a new set of nuclear PDFs with impact parameter dependence \cite{here:helenius}. In the absence of precise impact parameter dependent data, this new set, an essential tool to make phenomenological predictions, is based on previous fits with the b-dependence determined from their A-dependence.

More differential observables, especially involving correlations, are very promising in the quest for the saturation boundary in nuclear collisions. Generically, from a theoretical point of view, a n-particle differential spectrum needs of $2n$-point correlators. So, on top of the dipole scattering amplitude (related to the two-point correlator) extensively used in phenomenology, higher order terms would be needed. At present, this is one of the main topics of interest in the theory of the CGC, mainly motivated by the recent RHIC data on azimuthal 2-particle correlations at large rapidity in deuteron-gold collisions. Four talks in the conference where related with these n-point correlators: Jamal Jalilian Marian presented the evolution of a quadrupole from JIMWKL \cite{here:jalilian}; and both Dionisis Triantafyllopoulos (analytically) and Tuomas Lappi (numerically) presented an useful way to reduce these higher-order correlators to the better-known dipole case \cite{here:triantafyllopoulos}\cite{here:lappi}. Observable consequences where studied, in particular the description of the back-to-back azimuthal correlations of particles at large rapidity in d-Au at RHIC in talks by Heikki M\"antysaari \cite{here:mantysaari} and Amir Rezaeian \cite{here:rezaeian}. 

These studies, once the new data from the LHC pPb run will become available, will indeed constrain the nuclear partonic wave function in the small-$x$ region with unprecedented precision. One of the main question to be answer theoretically is how this initial state evolves into an equilibrated system, as the indications from hydrodynamical studies very strongly point to an equilibrated state formed quite early. This is an open question of fundamental interest as the only laboratory conditions to study the equilibration of a  non-abelian gauge theory are high energy nuclear collisions. New resummation techniques were presented by Fran\c cois Gelis which shown promising results, with only a scalar theory studied numerically at present \cite{here:gelis}.

Although {\it initial state effects} refers usually to the modification of the initial partonic distributions, it is often also used for the modification of the hadronization of the produced partons in the nuclear environment but in the absence of a hot medium. Two of these modifications were presented in the conference: on the one hand, Wei-Tian Deng presented effects of the modifications in the color flow of the decaying strings due to the presence of additional color sources in proton-nucleus collisions \cite{here:deny}; on the other hand, Xin-Nian Wang presented the corresponding results of energy loss by the cold nuclear matter \cite{here:wang}.

The next proton-lead run at the LHC, in January-February 2013, just before the Long Shutdown 1, will provide constraints to all these subjects. The sensitivity to the initial state effects will be much larger than even before and, importantly, the overlap in some regions of phase space with RHIC will allow to make stringent tests of the factorization hypothesis underlying the DGLAP analyses and the relevance of saturation effects in the partonic distributions.

\section{Quarkonia}
\label{section:quarkonia}

Quarkonia suppression/enhancement has been with us as one of the golden signals of QGP formation for the last 25 years. The high-quality data from the LHC and the systematics from RHIC and also SPS are providing strong constraints  on the underlying dynamics and the corresponding characterization of the medium properties. Lattice results presented in this conference show a significant progress on the question on what quarkonia bound states, if any, survive the transition region \cite{here:kaczmarek}. An updated MEM analysis of the charmonia states indicates that no signal for bound states above 1.46 $T_c$ is seen, this would indicate that the suppression of all charmonia states would be rather strong in the case of a phase transition. On the other hand, a different study, now for the upsilon states, indicates that they may be more resistant to dilution and survive up to relatively large temperatures. These results need to be taken with care, as further improvements could change the picture. Taken at face value, they would indicate that the mild suppression of $J/\Psi$ observed by ALICE could need of a recombination mechanism, while the suppression of the different $\Upsilon$ states measured by CMS would be compatible with sequential dissociation. More data from both LHC and RHIC, and, especially the benchmarking from pPb collisions will indeed help to elucidate the issue for which accurate QCD calculations would be also needed.    

From a phenomenological point of view, a powerful tool to implement the suppression dynamics into a calculation to be compared with experimental data is through effective theories and heavy quark potentials. Nice progress has been reached during the last years --- see the contributions to this conference by Alexander Rothkopf \cite{here:rothkopf} and Miguel Escobedo \cite{here:escobedo}.  The present uncertainties in the actual magnitude of the potential, including the essential role of its imaginary part, is still large for a sensitive comparison with experimental data. Two of these comparisons were presented at the conference by Georg Wolschin \cite{here:wolschin} and Che-Ming Ko \cite{here:ko}. 

One of the traditional difficulties on the interpretation of the quarkonia data has been the role of the so-called {\it cold nuclear matter effects}, which can be understood as a generic name for those effects affecting the production of quarkonia in nuclear collisions and that are not due to the presence of a hot medium. The usual way to pin down these effects, in the absence of a theoretically controlled implementation, is by measuring the suppression in pA collisions and extrapolating to AA by a Glauber model approach, including or not the modification of the nPDFs. Updated results of this approach were presented by Jean-Philippe Lansberg \cite{here:lansberg} and Ramona Vogt \cite{here:vogt}. These models offer a reasonable description of the data within the error bars around central rapidity with one fitting parameter for each collision energy. Further progress on the theoretical understanding on the dynamics of this suppression would be most welcome. In this respect, recent work presented at the conference by Fran\c cois Arleo, advocates for an energy loss origin of the suppression \cite{here:arleo} in line with some earlier approaches. A reasonable description of a large amount of pA/dAu data at different energies from SPS to RHIC is possible within this approach, interestingly, including large rapidities, normally difficult to reconcile with other models. The key point is a fractional energy loss $\Delta E/E\sim const.$ coming from initial-final state radiation interferences, neglected in previous calculations. 

\section{Electroweak probes}
\label{section:ew}

The reach of the TeV scale for the first time in nuclear collisions gives access to the production of weak bosons, a new tool which is only starting to show its potentialities. The almost null scattering strength of the weak bosons decaying into leptons within the created hot QCD matter make them scape the medium unaltered. In first approximation, then, no nuclear effects are expected and the nuclear modification ratio is unity if the Glauber model estimates of the number of collisions applies. This has been checked both at RHIC with photons and at the LHC with photons, $Z$- and $W$-bosons. Notice that this is an important check of the validity of the Glauber model, which is not a first-principle calculation, and on which a significant part of the experimental and phenomenological analyses rely. At present, the main interest of these probes is two-fold, on the one hand, they are expected to be excellent measurements for nuclear PDF analyses and to provide tight constrains to them; on the other hand,  $Z$+jet measurements have the advantage of a controlled calibration of the jet energy, essential for jet quenching analyses. Muon charge asymmetry from $W^\pm$ decays has been also presented in this conference by Bego\~na de la Cruz \cite{here:cruz}. The corresponding asymmetry in proton-(anti)proton collisions is included in global PDF analyses to constrain the up and down quark distributions. In the nuclear case,  in the absence of nuclear effects in the PDFs, the modification of the asymmetry is determined by the larger content on down quarks over up quarks with respect to the proton. The measured asymmetry, within the large error bars, agree with this isospin effect alone, and cannot distinguish the milder effects of the nPDFs. It is worth emphasizing, however, that the size of these modifications is determined by the relative corrections to up and down quarks in the nuclear PDFs which is {\it assumed} in all of the analyses to be the same at the initial scale. The experimental data provides support to this assumption and could be used in future analyses to constrain the corresponding uncertainty bands --- a study never carried out at present.

Photons produced at high transverse momentum agree also with the expectations from Glauber model plus nuclear PDFs, while those at smaller momentum provide a very useful tool to constrain  properties of the produced matter as the temperature. Charles Gale \cite{here:gale} made a review of the present situation and the uncertainties. Fluctuating initial conditions or viscosity  modify the yields of thermal photons, and a good description of the data is possible. The large $v_2$ measured, however,  would be more compatible with a hadron gas model, on the other hand incompatible with the yields. At present this is still a puzzle to be solved with more data and improved theoretical analyses. Dilepton yields, on the other hand, saw a nice progress from the theoretical side, with lattice and HTL calculations agreeing to a good degree of precision. Extending these calculations to the experimentally explorable regimes would lead to a more direct comparison between lattice/HTL theoretical results and data.

\section{Jet quenching --- including heavy quarks at high-$p_T$}
\label{section:jets}

Some of the most precious data from the LHC is again that from the high transverse momentum part of the spectrum. With reconstructed jets and inclusive particle suppression (both heavy and light hadrons) at RHIC and the LHC, the amount of information is becoming increasingly important, leading to strong constraints on the underlying dynamics.

\subsection{Phenomenology}

One of the surprising theoretical results in this conference is the excellent description of basically all available data on reconstructed jets by a simple model nicknamed {\it jet collimation} \cite{here:milhano}, presented by Guilherme Milhano. It basically assumes that the soft part of the partonic spectrum of the parton shower is taken out of the jet to large angles due to broadening (brownian motion in the transverse plane) with an additional energy loss due to BMDPS. The success of this simple model points to large angle energy flow as one of the key ingredients in the description of the jet data. At the same time it calls for new theoretical developments to provide solid grounds to a phenomenological model based on simple assumptions. In particular, issues as the role of the formation time would need to be taken into account --- also shown in the conference by Jorge Casalderrey-Solana are estimates based on Pythia where the typical formation time of most of the radiation in a shower is found to take place outside the medium \cite{here:casalderrey}.

Inclusive particle suppression at high transverse momentum is still an important source of experimental information. The LHC results show a suppression similar but slightly stronger than that at RHIC, with a positive slope of the $R_{AA}$ with increasing transverse momentum. Models tested at RHIC, based on radiative energy loss, are rather successful on the description of these data, as shown in the talks by Magdalena Djordjevic \cite{here:djordjevic}, William Horowitz \cite{here:horowitz} or Alessandro Buzzatti \cite{here:buzzatti}. This indicates that the dynamics underlying the suppression of high-$p_T$ particles is under rather good theoretical control, despite the fact that some uncertainties remain and need to be improved.

The new jets data from the LHC needs of new theory tools to make a full use of all their potentialities. In particular, the development of Monte Carlo (MC) tools was early recognized as an essential step towards a better interpretation of the data. Several different codes are being developed by different groups and using different approximations/assumptions. For the moment none of the implementations is completely satisfactory, sometimes lacking important ingredients in the phenomenological implementation, sometimes missing important ingredients in the underlying dynamical mechanism, sometimes both. Progress will need both rigorous analytical calculations (see next section) and information from experimental data. Here, the example of the {\it vacuum} Monte Carlo event generators is the one to follow: rigorous analytical calculations, including different types of resummations, constitute the kernel of the parton shower codes, on top of which phenomenologically developed mechanisms, as the hadronization model or the underlying event, are built based on experimental data. A Monte Carlo for jet quenching in the nucleus-nucleus counterpart will most probably need a similar approach.  Progress presented at this conference by Liliana Apolinario \cite{here:apolinario}, Clint Young \cite{here:young} and Thorsten Renk \cite{here:renk} is promising and helps to understand some of the properties or systematics of the experimental data. 

Transport models, including inelastic interactions, most frequently with LPM suppressions, were also presented in the conference. Denes Molnar \cite{here:molnar} presented results for $R_{AA}$ and $v_2$ of inclusive $\pi^0$; Berndt Muller presented results for the dijet asymmetry using the VNI/BMS model \cite{here:muller}; Jan Uphoff presented results for the heavy quarks both at RHIC and the LHC within the BAMPS model \cite{here:uphoff}; Min He presented also results for heavy quarks \cite{here:he}; as well as Marco Monteno, within a Langevin diffusion approach \cite{here:monteno}. 

New ideas in which gluon damping reduces the amount of radiative energy loss, and, at the same time, may make heavy and light quarks to lose energy more similarly were presented by Marcus Bluhm \cite{here:bluhm} and the corresponding phenomenological implementation by Pol B. Gossiaux \cite{here:gossiaux}. The proposal is based on a calculation in QED which is assumed to be valid for QCD. This is a reasonable proposal which is worth studying in more detail. Another new idea, presented by Rainer Fries \cite{here:fries} is the search for a quark $\longrightarrow$ photon conversion by triggering in a high-$p_T$ jet.

\subsection{New theoretical developments}

As stressed in the previous section, progress in the understanding and the calibration of the jet quenching as a probe for the medium needs rigorous analytical calculations on the modification of a parton shower in the presence of a medium. The phenomenology at RHIC was dominated by energy-loss effects which are not very sensitive to details of the underlying dynamics as, e.g. the relation of broadening and energy loss. Jet reconstruction and suppression, on the other hand, is mainly driven  by broadening while energy loss is also accessible by studying some of the jet substructure as, e.g. the fragmentation function. The recent data from the LHC only strengthens the need of such good theoretical control. 

For the energy loss of the leading particle, a successful mechanism based on medium-induced gluon radiation by a single emitter, iterated with some assumptions, e.g. the independent gluon emission approximation, is the most standard approach at RHIC. A complete parton shower needs, on the other hand, a good control on the multi-gluon emission probabilities. In the vacuum, this is a well-known procedure, in which ordering variables exist and the role of color coherence has been identified to be of great relevance. For the medium, the role of color coherence among different emitters has only started to be studied in the last couple of years and interesting consequences have arisen --- see the excellent review by Konrad Tywoniuk in these proceedings \cite{here:tywoniuk}. In these first studies, the simpler case of a $q\bar q$ antenna emitting a soft gluon has been considered. This allows to separate the first emission vertex from the soft gluons and to arrive to a simple physical picture in which the color coherence of the emitters play an essential role. A decoherence parameter, given by the dipole cross section, controls to what extent the quark and the antiquark radiate independently. When the typical scale of the medium is larger than the dipole size, the emission is coherent, and the BDMPS/GLV (independent) spectrum is suppressed. The opposite limit appears when quark and the antiquark rotate in color due to interactions with the medium and the color coherence is lost. In this case they are  independent emitters, a property that has been exploited by Yacine Mehtar-Tani and collaborators \cite{here:mehtar} to compute a generating function resumming all multiple medium-induced emissions enhanced by powers of $(\alpha_s L/t_{\rm form})^n$. To my knowledge this is the first time multi gluon resummation techniques are rigorously implemented in jet quenching calculations. In particular it provides support to the independent gluon radiation picture used in the literature so far. 

Color coherence  in the t-channel, where interferences between the initial- and final-state radiation can be studied, was also presented in the conference by Mauricio Mart\'\i nez \cite{here:martinez}. In this case, the interference pattern is modified with respect to the antenna (s-channel), technically this is due to some virtual corrections absent in the first case. Suppression of initial state radiation has been proposed, which may lead to a redefinition, in the nuclear case, of the resummations needed in some observables, leading to a more differential measurement of saturation-like effects.

In the non-perturbative side, modifying the color structure of the partonic jet shower due to interactions with the medium, may soften the fragmentation functions non-negligibly. The color flow in the vacuum parton shower follows a well defined pattern, better visualized in the large-$N_c$ limit. Color exchanges with the medium could lead to a modification of this color flow, resulting, e.g., in breaking strings stretching from the leading parton to the medium (at rest), in contrast with the much shorter strings stretching from the leading parton to the emitting gluon in the vacuum. These effects were presented by Andrea Beraudo \cite{here:beraudo}. 

As stated above, these new developments are in the right path in the quest for a well-defined and rigorous medium-modified parton shower algorithm, including modified splitting functions, coherence effects and color flows. On the other hand, in order to compare the outcome of the future phenomenological analyses with QCD, calculations of the transport coefficient $\hat q$ are also needed. Very nice progress has been reported in this meeting by Michael Benzke \cite{here:benzke} in which a gauge-invariant definition of the transport coefficient was presented and by Abhijit Majumder \cite{here:majumder} who presented the corresponding first lattice calculation. Although still with several limitations it looks as an extremely useful new line of developments in the field.   

\section{The AdS/CFT connection}

It has been a while since the first applications of the AdS/CFT correspondence with the experimental findings in high-energy nuclear collisions was performed. This fruitful relationship has developed an own line of research with interesting connections with other fields, also outside Particle Physics. The AdS/CFT correspondence is still our best theoretical tool to access some of the properties of the strongly interacting systems. In QCD this is especially true for dynamical quantities as the ones extracted from experimental data, e.g. the transport coefficient or the viscosity of the hot matter, where lattice calculations are very hard. At the same time, the field is reaching maturity in what concerns the {\it relation with reality}, {\it reality} here meaning QCD lattice results or experimental data. 

One of the traditional playgrounds for the AdS/CFT correspondence is the jet quenching phenomenology. Some new ideas were presented in the conference. Krishna Rajagopal \cite{here:rajagopal} made the analogy of the effects in a beam of gluons traversing a hot medium with the findings in experimental data --- in a strongly coupled medium, the beam of gluons would be attenuated in energy without spreading in angle, and without modifying the fragmentation function. Interestingly, jets are also advocated to be able to provide information about the relevance of a quasiparticle description of the plasma by the different broadening distributions in strongly-coupling vs weak-coupling systems. Mindaugas Lekaveckas \cite{here:lekaveckas} presented computations of the drag force in the background of two colliding shock waves to model the pre-equilibrium stage of the collision, finding, somehow surprisingly, that the drag force turned out to be somehow smaller than if the plasma were instantaneously in equilibrium, meaning no 'extra' energy loss from being far from equilibrium. Derek Teaney \cite{here:teaney} was in charge of reviewing the last developments in the AdS/CFT correspondence applied to heavy ion collisions. Among the different calculations, there are interesting progress on the spectral functions, with recent lattice data which somehow turns out to be in between the strong- and weak-coupling expectations. In the same talk, the difficulties with the energy loss of light quarks was also discussed, a topic which was also discussed in the parallel session by Andrej Ficnar \cite{here:ficnar}.

\section{Conclusions and (mainly) future prospects of hard probes in nuclear collisions}

Hard probes is one of the pillars of heavy-ion collisions phenomenology, even more so since the beginning of the LHC. The impressive new data becoming available on reconstructed jets, heavy quarks, different quarkonia states, new access to electroweak probes, etc., require new more and more precise theoretical tools. At the same time, the hardest scales allow the formalisms to be under better control and to perform calculations in QCD with a lower degree of modeling. The future of the hard probe studies relies on more and more precise first-principle calculations, better adapted to compare directly with data. 

We are still at the beginning. More progress is ahead which requires hard theoretical work among other topics on

\begin{itemize}

\item {\it The role of initial conditions}. Saturation calculations based on the Color Glass Condensate approach, and particularly on the numerical solution of the NLO BK equation have seen an enormous progress in the last couple of years. Comparison with experimental data is more and more accurate --- {\it with solutions of the QCD evolution equations}. This is the right path to follow for the phenomenology and the new data expected for the next proton-lead run will be essential in our understanding of the relevance of saturation of partonic densities.  On the DGLAP side, the formalism has been set long ago and is only waiting for experimental data to proof or disproof  the ability of the linear equations and the factorization hypothesis to deal with the nuclear case in the new kinematical regimes of small-$x$. The reach of the TeV energies allows also to probe de nPDFs with new and more precise tools as the $W/Z$ bosons. The partial overlap in $(x,Q^2)$ phase space with RHIC will provide the needed constraints in all cases. The impact parameter dependence both for the linear and the non-linear dynamics is available but needs of some assumptions which should become more constrained by the new data. Notice that while the nPDFs in DGLAP analyses are the essential ingredients for a clean phenomenological interpretation of the signals to characterize the hot medium, the CGC is to date the best, we could say only, general framework in which a general understanding of the whole collision dynamics, from the initial state to the thermalization, can be realized. Each approach have its own range of validity which can only be pinned down by experimental data.

\item {\it Understanding the intriguing quarkonia data}. Is there a simple explanation for the quarkonia data given the fact that some of the systematics are quite simple --- centrality dependences, different quarkonia states suppression... i) If sequential suppression is the right explanation for the upsilon states then a more direct comparison with lattice calculations should be possible, for that the theoretical uncertainties in the calculations would need to be reduced. ii) If recombination is at work for the charmonia, better control on the theoretical implementations would be most welcome. iii) All quarkonia data needs of a benchmarking proton-nucleus data for which a real theoretical control has been historically a serious drawback, new proposals presented in the conference are extremely important if they are proven to be right. iv) Effective theories and potential models, very useful for phenomenology, would need a better control --- also here, progress has been substantial, though. 

\item {\it A better theoretical control on the in-medium parton shower}. The new developments, triggered by the first published data on reconstructed jets, provide a more solid fundamentals to the implementation of a parton shower algorithm in the presence of a medium. The role of interferences, color flow, etc, essential in any description of the vacuum parton shower, is being clarified, and some non-expected features appear. These solid analytical derivations, possibly implemented in Monte Carlo codes,  need to be performed for a consistent and rigorous study of the experimental data. 

\item {\it Monte Carlo codes for jet quenching studies}. The need of Monte Carlo codes for the phenomenology of jet quenching as well as for the experimental studies is very clear. During the conference a discussion session on this topic was scheduled where several different points of view where raised. In my opinion, the development of MC codes would need of an iterative theory/experiment approach, similar to the one for proton-proton codes, in which the hard-core, the underlying perturbative physics, is under good theoretical control from analytical calculations in different limiting cases. On top of this hard-core, other effects would need to be built using experimental information and/or reasonable modeling. The interest of Monte Carlo tools is that some of the mechanisms, difficult to implement in analytical calculations, as energy momentum conservation, or different medium geometries, may be straightforward --- here {\it geometry} is used generically, it includes hydrodynamic profiles, for example. This provides MC approaches a great versatility to study the systematics in data, to find corrections, etc. 

\end{itemize}

\section*{Acknowledgments}
I would like to thank the organizers of Hard Probes 2012 for the excellent work performed in putting together an exciting program and for giving me the opportunity to summarize its theory part. 
This work is supported by European Research Council grant HotLHC ERC-2001-StG-279579, by Ministerio de Ciencia e Innovaci\'on of Spain under grant FPA2009-06867-E and by Xunta de Galicia.

\bibliographystyle{elsarticle-num}



\end{document}